# Intra-site Level Cultural Heritage Documentation: Combination of Survey, Modeling and Imagery Data in a Web Information System

Élise Meyer [1], Pierre Grussenmeyer [2], Jean-Pierre Perrin [1], Anne Durand [3], Pierre Drap [3]

[1] MAP-CRAI UMR CNRS-MCC 694, Nancy School of Architecture, Nancy, France
(meyer, perrin)@crai.map.archi.fr

[2] MAP-PAGE UMR CNRS-MCC 694, INSA Strasbourg Graduate School of Science and Technology, Strasbourg, France
pierre.grussenmeyer@insa-strasbourg.fr

[3] MAP-GAMSAU UMR CNRS-MCC 694, Marseille School of Architecture, Marseille, France
(anne.durand, pierre.drap)@gamsau.map.archi.fr

**Abstract**
*Cultural heritage documentation induces the use of computerized techniques to manage and preserve the information produced. Geographical information systems have proved their potentialities in this scope, but they are not always adapted for the management of features at the scale of a particular archaeological site. Moreover, computer applications in archaeology are often technology driven and software constrained. Thus, we propose a tool that tries to avoid these difficulties. We are developing an information system that works over the Internet and that is joined with a web site. Aims are to assist the work of archaeological sites managers and to be a documentation tool about these sites, dedicated to everyone. We devote therefore our system both to the professionals who are in charge of the site, and to the general public who visits it or who wants to have information on it. The system permits to do exploratory analyses of the data, especially at spatial and temporal levels. We propose to record metadata about the archaeological features in XML and to access these features through interactive 2D and 3D representations, and through queries systems (keywords and images). The 2D images, photos, or vectors are generated in SVG, while 3D models are generated in X3D. Archaeological features are also automatically integrated in a MySQL database. The web site is an exchange platform with the information system and is written in PHP. Our first application case is the medieval castle of Vianden, Luxembourg.*

H.2.8 [Database Management]: Database Applications ; H.5.1 [Information Interfaces et Presentation]: Multimedia Information Systems ; J.2 [Physical Sciences and Engineering]: Archaeology

## 1. Introduction

Cultural heritage documentation induces the use of computerized techniques to manage and preserve the information produced. In the archaeological domain particularly, data computerization gives solutions to specific problems in allowing inventory actions to save, present or interpret the features. Archaeology is an erudition discipline where the knowledge grows up in necessarily referencing the precious documents already gathered. Computer science has fast appeared as a very convenient way to manage this information, which the development of the field is enriching at great speed in an inflation quite difficult to master. It is required to develop systems allowing to create relationships between the different types of information generated by the working of an archeological site. We have chosen to work at the scale of a site, because there are more completed projects at regional or national levels. In addition, these projects are mostly Geographic Information Systems, whereas we are creating an Information System permitting to manage very different types of data (not only geographical) what is more uncommon. We dedicate besides our system both to the professionals who are in charge of the site and to the general public who visits it or who wants to have information on it. Hence, we are using web techniques and languages to develop our web information system, in order to be independent of any software and to allow maximum accessibility and adaptability to the needs of the diverse actors using the tool. Our first application case is the medieval castle of Vianden, located in Luxembourg.

The system notably combines survey, modeling and imagery data, and our purpose is to highlight up to what point can such a system offer new possibilities for the management and the documentation of the data of an archaeological site. First of all, we will present the objectives of the project and the methods chosen to reach our aims, according to the state of the art in the domain. Afterwards, the tool that we are developing will be presented. We will describe how the different data types are recorded in the system. Then, we will show the access possibilities to this data: through queries and through 2D and 3D interfaces. Finally, we will explain on the prototype of the Vianden castle site the way to create original 3D models and synthesis plans, and the means to update and revise the data for the experts working on the site.

## 2. Objectives of the project

To introduce the aims of the project presented in this paper, we will give the principal conclusions of the bibliographical study that has been made in the beginning of our work. So, we will correlate the fixed objectives with the lacks and needs identified during this study.

A good overview on computer applications in archaeology has been given by J. D. Richards [Ric98]. This paper reflects that "although archaeologists have been quick to apply the latest technology, in most cases the technological driving force has been outside the discipline" [Ric98]. What means that the use of computer science in the archaeological domain is often driven by software offers rather than by archaeological questions. This is a problem to which we will try to propose solutions.

Concerning databases, Richards write that the description of a document that is recorded in a database is at least as rich as the report from traditional publications. The recording of metadata (data about data) is something common nowadays, notably with the format XML (Extensible Markup Language, standard language of the W3C) which is dedicated to the formalization of such information and which allows polymorphism. This is the format that we have adopted to record metadata about the archaeological features that our system permits to manage. The idea is that an excavation archive can be viewed as a hyper document with texts and images bounded by internal links and allowing readers to follow different paths to retrieve information through the report [Rya95]. And "if such documents can be made publicly accessible, over the Internet, for example, then they begin to blur the distinction between archive and publication". So we have chosen to develop a free Information System that works over the Internet. An example of integrated computerized field projects linking basic finds, plans and context data recording in the field and operated using GIS mapping tools is related in [Pow91], and examples of Internet databases applications are given in [CFR03] and [Ric04].

A drawback of projects carried out currently, according to Richards and as already said in introduction, is the fact that a large proportion of the literature until now has been concerned with the establishment of databases of archaeological sites and monuments at regional and national levels for cultural resource management purposes. They are few projects concerning data management at a site level, the data recorded being obligatory dissimilar at this scale than at a bigger. Consequently, our project is devoted to the management of data generated by the working of a particular site (and not of a group of sites).

Regarding Geographical Information Systems, they have been developed to create relationships between data and to analyze spatial information recorded in databases. In archaeology, the principal applications of GIS are either heritage management (monitoring of known sites or identification of new ones) or explanatory framework (site catchments or view shed analysis). A great quantity of examples are cited in Richards' paper and his conclusion is the following: "There has been a lack of projects that have made effective use of GIS at the intra-site level; the projects on an Iberian cemetery [QBB95], Roman Iron Age sites in the Assendelvers Polders [Mef95], and the Romano-British settlement at Shepton Mallet [BCE*95] are rare exceptions." [Ric98]. For that reason, the project that we are developing concentrates on this lack. We generalize the notion of Geographical Information System in saying only Information System to describe our work. In fact, the types of data that are managed are not only geographical data (maps, vectors) but also archaeological, historical, topographical, architectural, geological, environmental… An information system must permit to carry out a real multidisciplinary synthesis of all resources of the database. For archaeological data especially, the creation of an information system can lead to achieve:

- to treat graphically several information derived from very different kinds of surveys, because a selective superposition could be a precious help for the interpretation;
- to combine elements selected in diverse graphs for the carrying out of visualizations in a synthesis plan;
- to present images and their connections with the concerned texts from the database, to lead to a complex system in which the examination of texts and images would be possible simultaneously.

Especially for our project, the aim is to create a tool permitting to manage data generated by the working of an archaeological site. The term management comprises the gathering and description (metadata) of all the documents (photographs, plans, drawings, models…) already created or that will be done during the further exploitation of the site, and the construction of relationships and links between them.

To continue, visualization of archaeological information is one of the most exciting ways in which computer technology can be employed in archaeology. This word is taken for any exploration and reproduction of data by graphical means. The use of this technique allows visual interpretation of data through representation, modeling, display of solids, surfaces, properties or animation, what is rarely amenable to traditional paper publication [Ric98]. Until now, the most 3D models are intended for heritage center and museums displays, rarely are some available on-line over the Internet. An impressive and popular publication of visualizations of international important sites has been edited by [FS97]. Also virtual reality with fully immersion has a great potential as a medium for interpretation and communication to the general public [GE04]. An example of web-based visualization in VRML that allow to explore an archaeological landscape (large scale) is given in [GG96], and explanations about the use of the SVG format are to read in [Wri06]. One of the principal inconvenience of the types of 3D models used in archaeology until now is that these models are blank. In fact, they only serve for visualization needs and they don't give any other information. Nowadays, 3D models can serve as research interface to access different kinds of information, notably in coupling them with web procedures (scripts). Our project is carrying out this way: we are producing interactive plans in SVG and models in X3D that work like web interfaces to access the database data.

Finally, when spoken about communication in the domain that interests us here, it is often heard as publication. More precisely, significant developments regarding communication currently have appeared with new forms of electronic publications. Electronic publication allows the distinction between traditional archive and hard copy report to blurred, with supporting data made accessible for the first time [Rya95]. From another source, there are advantages through multimedia and accessibility of new forms of data, particularly drawings, plans, video, and photographs [RS94][Smi92]. [McA95] note that doubts have been expressed about the speed of adoption because of resistance from traditional publishers. We can say now that these doubts were well-founded because there are not yet a great quantity of electronic publications in archaeology, especially available over the Internet. However, one of the best examples of on-line publication (peer-reviewed journal of record) is *Internet Archaeology*, an international electronic journal project set up with funding from the United Kingdom's Higher Education and Further Funding Council (HEFCE) as part of their eLib (Electronic Libraries) program [HRR95]. This publication doesn't contain any other material than textual documents. According to Richards, "undoubtedly the major growth area of the second half of the 1990s has been that of archaeology on the Internet, particularly on the World Wide Web" [Ric98]. This is even more true today, the web provides a tremendous opportunity to link distributed resources and to make unpublished material widely available (remarkably uncommon material like detailed fieldwork data, quantities of photos and archive drawings, vectorial plans or 3D models). The traditional division between publication and archive could thus be removed, even if there is still a big challenge to control the way in which the Internet is used (for the discoveries, quality controls or copyrights). From our side, the way we perceive the term communication is more complete than just the publication. Aim of the web site including the information system, is to assist the digital archiving of the documents, their inquiry and their processing by everyone, both the professionals (archaeologists, surveyor, architects, etc.) and the general public. Different types of access to the data are available depending on the user of the system. Representations adapted to museum displays (public attractive) have been done as well as interfaces permitting to update the data directly from the 3D models (for instance) for the needs of the site managers. This system works over the Internet to allow accessibility and simplicity for all the users, and above all to be free from any software. As a conclusion, Richards said that "in all areas of computer applications in archaeology, the discipline has been technology driven and software constrained. Rarely has the use of computers in archaeology been led by archaeological theory, although in specific fields, such as GIS, it can be demonstrated that computers have advanced archaeological knowledge." [Ric98]. Thus we hope the information system we are developing will also serve archaeological knowledge in proposing an other type of communication and sharing of the information generated by an archaeological site.

### 3. Implementation of a web information system dedicated to archaeological intra-site features documentation

According to the objectives explained before, we will now present the way the project has been developed to reach these aims in the best possible way.

To begin with, it is relevant to point out the fact that the computational base of the information system carried out comes from projects done to integrate photogrammetric data and archaeological knowledge on the web (ISA-PX "Information System for Archaeology using Photogrammetry and XML"). These projects are parts of research of P. Drap and his team [DG00] [DDS*05].

In fact, a certain number of the computer formats (XML, VRML) used in the ISA-PX system were adapted to our needs. The laboratory of P. Drap being partner with us, it has been possible to base our project on the computer developments already done. We have adapted the existing system afterwards to our particular case, notably to allow the management of different types of data (not only data coming from photogrammetric surveys), and coupled with a web site to permit simple data access by everyone. Our system has been named SIA (Archaeological Information System).

### 3.1. Database management system

The types of data that the system allows to manage are:
- temporal data (description of historical periods)
- spatial data (description of places of the archaeological site)
- different sorts of plans that have been digitized (axonometries, maps, sections, plans, elevations, excavation profiles and plans)
- digital photos or ancient photos digitized
- scanned drawings
- scanned texts
- vectorial plans (generated in SVG)
- 3D models (generated in X3D)

These data are recorded both as XML files and in tables of a MySQL database. These two record possibilities were already available in the ISA-PX system, to obtain standardized data formulized in XML (for simple information exchanges) and a classic form of data searchable through SQL queries in MySQL. More precisely, it is metadata about the before quoted data that is recorded (for instance the provenance, author, subject, coordinates… of a photo). Figure 1 sums up the process to fill the XML and MySQL databases (both are filled simultaneously).

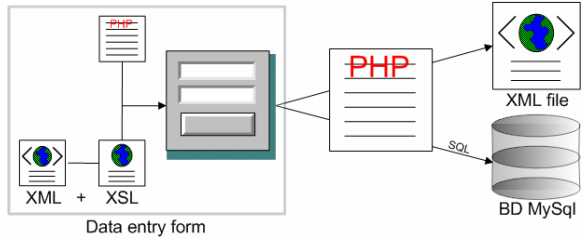

**Figure 1:** *Filling of the databases (computer behavior).*

After having integrated a first time the corpus of each data type in the form of an XML file generated by the system (data entry form to give the metadata structure), all the metadata is recorded through data entry forms like in Figure 1. The data itself is attached thanks to URL links.

Figure 2 gives an example of the HTML representation of an XML photo file (data and metadata) thanks to an external XSL document. When the user clicks on the miniature of the photo, he have access to the original photo.

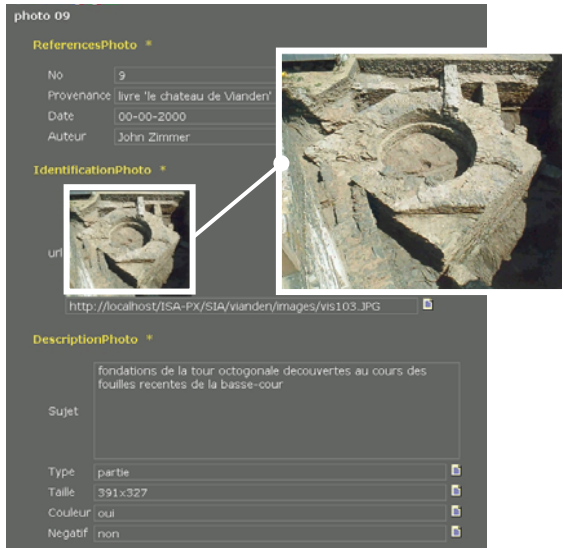

**Figure 2:** *HTML representation of the XML data and metadata of a photo thanks to an external XSL document.*

### 3.2. Accesses to the data in the system SIA

Figure 3 illustrates the computer behavior of the platform that has been developed. The initial information system ISA-PX has been totally included and adapted to the web interface that has been created to form the SIA system, which permits a user friendly and insightful access to the archaeological data recorded.

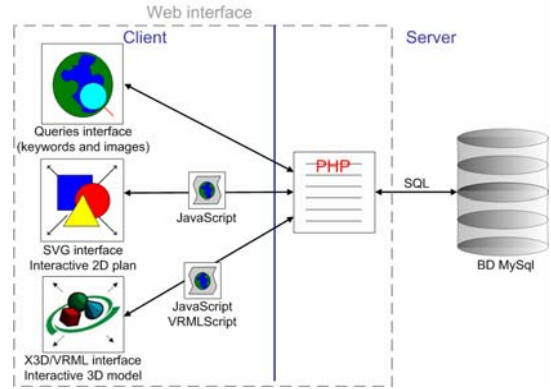

**Figure 3:** *Types of data accesses in the SIA Information System (computer behavior).*

In parallel to these accesses, two menus are available in the web site to retrieve documents in covering the history of the site and in visiting its places. To sum up, the documents inquiry in the system SIA is schematized in Figure 4.

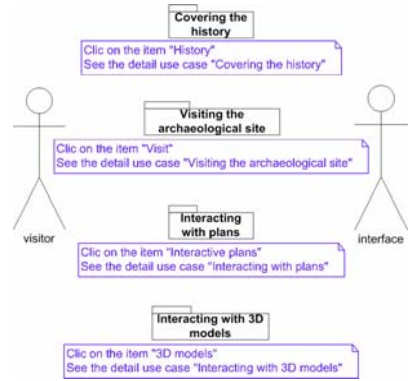

**Figure 4:** *User's path for the documents inquiry (UML).*

Each use case (presented as a folder here) is also detailed in an other UML schema showing precisely the different operations to do for an efficient exploitation of the SIA information system. For instance, Figure 5 shows the detailed use case "Interacting with plans".

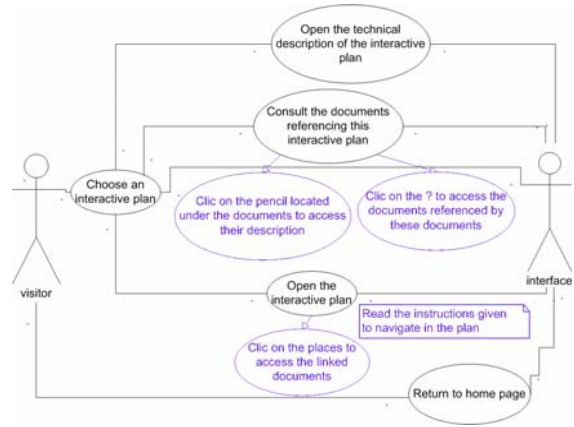

**Figure 5:** *SIA use case "Interacting with plans" (UML).*

All the UML schemas done to help the users of the system (visitors and expert users) are available in the web site in an item entitled "help and users' paths". Likewise, the queries interfaces (by keywords and images) are explained thanks to UML schemas, to allow the user to find a document as fast as possible if he has particular criteria. The keywords search engine is multi-criteria, what means that the user has choice between different words, he doesn't give the words himself.

### 3.3. Example of the Vianden castle site

The place of interest on which we have done our first experimentations is the medieval castle of Vianden located in North-East Luxembourg. This archaeological site has a very long and interesting history, during which a lot of documents have been created and hold. Then, this site was very interesting for us to test our system on a real case, notably because we have had access to historical models of the castle made by the MAP-CRAI laboratory of Nancy. So we have collected in the SIA system a lot of data produced by the conservators of the castle (plans, photos, excavations profiles…), along with the MAP-CRAI 3D models (converted from the Maya® format into VRML) and with interactive plans that we have created in SVG thanks to the software Adobe Illustrator®.

All these documents gathered and created have been registered in the database (themselves and the metadata attached) in using the procedure explained in Figure 1.

The principles of accesses to the data are explained in Figure 3. In a more detailed way, Figure 6 shows the example of Vianden site.

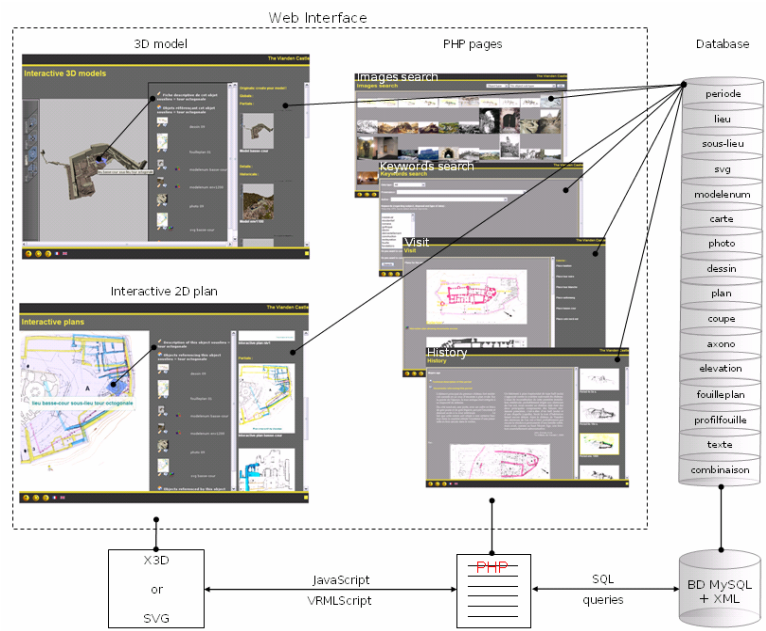

**Figure 6:** *SIA information system in the case of the Vianden castle.*

**Creating models and plans.** In addition to the data working possibilities explained before, the user of the SIA system can create his own models and composition plans thanks to multi-criteria data entry forms. In the web pages, he has the choice to select for example one or several places and one or several historical phases for which he wants to create "on the fly" the 3D model to see the evolutions in time of the castle. The procedure is to see in Figure 7. The resulting model presents different parts of the castle (yard, chapel, hall) in two different periods (year 1100 in yellow and 1150 in pink). We can see here the architectural changes that have been done during the 50 years considered. The same process is available to create synthesis plans or photo-montage allowing for instance to superimpose the physionomy of the castle today with its former aspects.

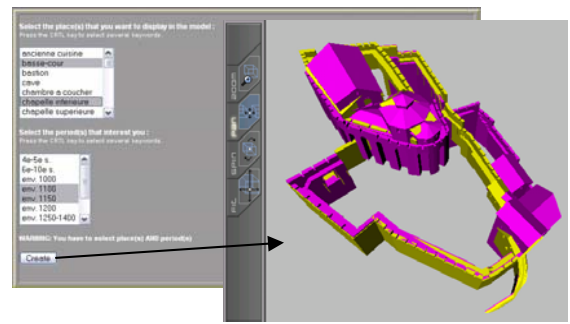

**Figure 7:** *Process to create on the fly 3D models.*

**Updating the data.** To look at the data is the first step in the analysis of a dataset. To go further on, the archaeologist (or any expert that is logged on) needs to entry/edit them.

To update the metadata:
- the metadata corpus can be used to correct some basic errors (misspelling, simple inconsistencies, etc.),
- through the graphic interfaces in SVG or VRML, the expert user can directly modify the selected object,
- through different types of research modes (by object type, by location, by epoch, …), the user can straight access to the data and can edit it for modifications.

For the revising, the conceptual model used to describe the objects can change during the time of the study, according for example to new archaeological knowledge. The expert user can modify accordingly the tree structure of the dataset describing the object model.

## 4. Conclusion and future work

After having set our work objectives in the state of the art of computer applications in archaeology, we have introduced the Information System that we have developed. The web site and the underlying information system allow to record, make use and represent the data of any archaeological site. The SIA system has been made to search solutions to help archaeologists in their tasks at an intra-site level and to avoid that they are software-driven. The full XML choice for textual and graphical representations permits relevant interactions. The use of 2D vectorial graphics and 3D models as user-interfaces to the data link purely documentary data and metadata to geometric representations. We connect very different types of data to emphasize new research possibilities and new information exchanges between many sites to be able to draw conclusions by crosschecking. Moreover, the data are available through the Internet what allows us to work in the direction of communicating them in an innovative and interactive way. Experiments will be carried out soon on a gallo-roman site to highlight the subsisting problems (integration of new data types…) and to test the clarity of the help files created (UML schemas) in order to know if the system is usable by everybody.

Final aim is to create a simple and everywhere accessible tool for all the archaeological sites managers, who wish to be able both to exploit the quantity of data produced and to represent them, in order to make use of this archaeological information system as a virtual storefront for the communication and the e-publication of their findings.